\documentclass[epj]{svjour}
\usepackage{graphics}
\usepackage{graphicx}
\usepackage{amsmath,amsfonts,amssymb}
\usepackage{mathrsfs}
\usepackage{extarrows}
\usepackage{color}
\begin{document}
\title{Analytical controllability of deterministic scale-free networks \\
and Cayley trees}

\author{Ming Xu\inst{1,2} \and 
Chuan-Yun Xu\inst{1} \and 
Huan Wang\inst{3} \and 
Cong-Zheng Deng\inst{2} \and 
Ke-Fei Cao\inst{1,}
\thanks{e-mail: kfcao163@163.com}%
}                     
%
%
\institute{
Center for Nonlinear Complex Systems, 
Department of Physics, 
School of Physics Science and Technology, 
Yunnan University, 
Kunming, Yunnan 650091, P.R. China
 \and 
School of Mathematical Sciences, 
Kaili University, 
Kaili, Guizhou 556011, P.R. China
 \and 
School of Computer Science and Technology, 
Baoji University of Arts and Sciences, 
Baoji, Shaanxi 721016, P.R. China
}
%
\date{{\color{blue}Eur. Phys. J. B} (2015) {\bf 88}(7): 168\\
Received 3 November 2014 / Received in final form 23 March 2015 / 
Published online 1 July 2015\\
The final publication is available at Springer via 
{\color{blue}http://dx.doi.org/10.1140/epjb/e2015-60230-3}}
%
\abstract{
According to the exact controllability theory, the controllability is investigated analytically for two typical types of self-similar bipartite networks, i.e., the classic deterministic scale-free networks and Cayley trees. Due to their self-similarity, the analytical results of the exact controllability are obtained, and the minimum sets of driver nodes (drivers) are also identified by elementary transformations on adjacency matrices. For these two types of undirected networks, no matter their links are unweighted or (nonzero) weighted, the controllability of networks and the configuration of drivers remain the same, showing a robustness to the link weights. These results have implications for the control of real networked systems with self-similarity.
\PACS{
      {89.75.-k}{Complex systems}   \and
      {02.30.Yy}{Control theory}
     } 
} 
\maketitle
\section{Introduction}

The control of complex networks is of paramount importance in network science and engineering, which has received extensive attention in the past decade or so~\cite{S2001n,LH2007pre,M2008ctcn,LSB2011n,YZD2013nc,LYF2014epl,L2014nsr}. In control theory, the controllability property plays a pivotal role in many control problems. Generally speaking, a dynamical system is controllable if it can be driven from any initial state to any desired final state in finite time by a suitable choice of inputs~\cite{K1963jsiama,SL1991anc}. The theory of controllability is mathematically sound, and has been widely applied to engineering. However, it is difficult to apply the traditional controllability theory directly to complex networks. Here, the first challenge faced is how to find the minimum set of driver nodes (i.e., drivers) needed to fully control the whole network~\cite{LH2007pre}, which is a computationally prohibitive task for large networks by the traditional Kalman's controllability rank condition~\cite{K1963jsiama,R1996lst}. In recent years, controllability has become a hot research topic in the field of complex networks~\cite{LSB2011n,YZD2013nc,LYF2014epl,NP2013tac,RR2013cn,NWZ2014po,C2014ijcas,JP2014sr,YZW2014njp}. Liu et al. introduced a paradigm to study the structural controllability of an arbitrary complex directed network; this is an important progress in identifying a minimum set of drivers~\cite{LSB2011n}. Another significant recent contribution was made by Yuan et al.~\cite{YZD2013nc}; according to the Popov-Belevitch-Hautus (PBH) rank condition~\cite{R1996lst,H1969pknawa,S1998mctdfds}, they developed an exact controllability framework to determine the minimum set of drivers from the maximum multiplicity of eigenvalues of the network matrix by elementary column transformation~\cite{YZD2013nc}. With this framework, the exact controllability has been studied for many networks such as Erd\H{o}s-R\'{e}nyi random networks, scale-free networks, small-world networks, simple regular networks, and some real networks~\cite{YZD2013nc}. Furthermore, some analytical results about the minimum number of drivers, which are determined by the algebraic multiplicity of the eigenvalue $0$ of the adjacency matrix, are obtained in reference~\cite{LYF2014epl} for three typical regular fractal networks including the modified $(1,2)$-tree network, the Peano network and the modified dual Sierpi\'{n}ski gaskets network. It is hard work to find analytical results for the minimum set of driver nodes in large networks, even in simple regular systems such as path and cycle graphs~\cite{PN2012tac}, grid graphs~\cite{NP2013tac}, and some regular fractal networks~\cite{LYF2014epl}.

In this paper, the exact controllability is explored for two typical types of self-similar bipartite networks including the classic deterministic scale-free network (DSFN)~\cite{BRV2001pa} and Cayley trees~\cite{CC1997m,JWZ2013jcp}. Here, by improving the framework in reference~\cite{YZD2013nc}, we obtain the analytical results for both the minimum number and the minimum set of drivers using elementary row and column transformations on adjacency matrices. Since scale-free networks and tree-like networks are abundant in nature and society, and Cayley trees can model the structure of dendrimers, a classic family of macromolecules; therefore, the analytical method used in this paper and the study of the controllability for these two types of networks are of theoretical and practical significance for networks with self-similarity.

This paper is arranged as follows. First, we briefly introduce the exact controllability theory and provide a sufficient condition for the expression (Eq.~(\ref{ND.max.N.A})) to determine the minimum number of drivers for undirected bipartite networks. Second, the self-similarity in these two types of networks allows us to find some properties of the coupling matrix. By properly using elementary transformations of a matrix, we can find the minimum set of drivers and obtain analytical results of the controllability in these networks. Finally, the distribution characteristics of drivers and a robustness of the controllability to the link weights are discussed.

\section{Exact controllability theory}

Consider a network with $N$ nodes described by the following system of linear ordinary differential equations~\cite{LSB2011n,YZD2013nc}: 
\begin{equation}
\label{linear ODEs}
\dot{\textbf{x}}=A\textbf{x}+B\textbf{u},
\end{equation}
where $\textbf{x}=(x_{1},x_{2},\dots,x_{N})^\text{T}$, $\textbf{u}=(u_{1},u_{2},\dots,u_{M})^\text{T}$, $A\in \mathbb{R}^{N\times N}$, and $B\in \mathbb{R}^{N\times M}$ denote, respectively, the vector of the states of $N$ nodes, the vector of $M$ controllers, the coupling matrix of the network, and the input (control) matrix. According to the PBH rank condition, system (\ref{linear ODEs}) is fully controllable if and only if 
\begin{equation}
\label{rank cI.A.B}
\text{rank}(cI_{N}-A,B)=N
\end{equation}
is satisfied for any complex number $c$, where $I_{N}\in \mathbb{R}^{N\times N}$ is the identity matrix. Thus, the minimum number $N_{D}$ of independent drivers required to control the whole network, which is defined by $N_{D}\equiv \min\{\text{rank}(B)\}$~\cite{LSB2011n}, can be deduced as~\cite{YZD2013nc}: 
\begin{equation}
\label{ND.max.mu}
N_{D}=\max\limits_{i}\{\mu(\lambda_{i})\},
\end{equation}
where $\mu(\lambda_{i})\equiv N-\text{rank}(\lambda_{i}I_{N}-A)$ stands for the geometric multiplicity of the eigenvalue $\lambda_{i}$ of matrix $A$. For a symmetric coupling matrix $A$, equation~(\ref{ND.max.mu}) can be written as follows~\cite{YZD2013nc}: 
\begin{equation}
\label{ND.max.delta}
N_{D}=\max\limits_{i}\{\delta(\lambda_{i})\},
\end{equation}
where $\delta(\lambda_{i})$ is the algebraic multiplicity of the eigenvalue $\lambda_{i}$, which is also the eigenvalue degeneracy of matrix $A$.

Especially, for a large sparse undirected network, in which the number of links is of the same order as the number of nodes in the limit of large $N$~\cite{N2004pre}, in the absence of self-loops or with a small fraction of self-loops, the maximum geometric (or algebraic) multiplicity of the coupling matrix occurs at the eigenvalue $\lambda=0$ with high probability~\cite{YZD2013nc,PP2002prvsp}, which yields a simplified expression for $N_{D}$~\cite{YZD2013nc}: 
\begin{equation}
\label{ND.max.N.A}
N_{D}=\max\{1,N-\text{rank}(A)\}.
\end{equation}
Since equation~(\ref{ND.max.N.A}) is not always true for sparse undirected networks, for example, in a ring network with $N$ nodes, whose eigenvalues are given as $\lambda_{i}=2\cos(2\pi(i-1)/N)$ ($i=1,2,\dots,N$), it is known that $N_{D}=2$~\cite{YZD2013nc}, whereas $\max\{1,N-\text{rank}(A)\}=\max\{1,0\}=1$ ($\neq N_{D}$) when $N$ is a prime number; hence, if a sufficient condition is given for equation~(\ref{ND.max.N.A}), it will be convenient to use equation~(\ref{ND.max.N.A}) for determining $N_{D}$.

As we have known, a graph (network) $\Gamma$ is bipartite if and only if for each eigenvalue $\lambda$ of $\Gamma$, $-\lambda$ is also an eigenvalue, with the same multiplicity~\cite{BH2012sg}. So, for an undirected bipartite network, if the following sufficient condition 
\begin{equation}
\label{suff.cond.}
\delta(0)\geqslant\frac{N}{3}
\end{equation}
is satisfied, then 
\begin{equation}
\label{ND.delta}
N_{D}=\max\limits_{i}\{\delta(\lambda_{i})\}=\delta(0)
\end{equation}
and equation~(\ref{ND.max.N.A}) hold.

How can we find the minimum set of drivers for an undirected network when equation~(\ref{ND.delta}) holds? According to the exact controllability theory~\cite{YZD2013nc}, the control matrix $B$ to ensure full control should meet the condition (Eq.~(\ref{rank cI.A.B})) by substituting $0$ for the complex number $c$, as follows: 
\begin{equation}
\label{rank A.B}
\text{rank}[-A,B]=N,
\end{equation}
which means that the $N$ rows of $[-A,B]$ is linearly independent. To find the drivers from $B$ to satisfy equation~(\ref{rank A.B}), we should select a maximal linearly independent group from all row vectors of $A$, and the redundant rows would correspond to the drivers. Note that the configuration of drivers is not unique because there are many possible choices of maximal linearly independent groups. Nevertheless, the minimum number of drivers is unchanged.

Elementary transformations of matrices play a crucial role in finding the drivers. There are three types of elementary row transformations of a matrix~\cite{W1934lm}: interchanging rows $i$ and $j$, denoted by $\mathscr{R}_{i}\leftrightarrow \mathscr{R}_{j}$; multiplying a row (say, the $j$th row) by a nonzero number $k$, denoted by $k\mathscr{R}_{j}$; and adding $k$ times the $j$th row to the $i$th row, denoted by $\mathscr{R}_{i}+k\mathscr{R}_{j}$, where $i\neq j$. Similarly, we can also define elementary column transformations by changing ``row'' into ``column''(the corresponding symbol is changed from ``$\mathscr{R}$'' to ``$\mathscr{C}$'', respectively). Note that the rank of a matrix remains unchanged by elementary matrix transformations. Furthermore, if matrix $A_{1}$ can be obtained from matrix $A_{2}$ by a series of elementary operations $\mathscr{R}_{i}+k\mathscr{R}_{j}$ ($i\neq j$), then row $i$ of $A_{2}$ is redundant $\Longleftrightarrow$ row $i$ of $A_{1}$ can become a redundant row.

According to reference~\cite{LSB2011n}, the controllability of a network can be defined by the ratio of $N_{D}$ to the network size $N$, i.e., 
\begin{equation}
\label{nD.def.}
n_{D}=\frac{N_{D}}{N}.
\end{equation}
In the following, we analytically derive $N_{D}$ and $n_{D}$ of two typical types of self-similar bipartite networks and consequently discuss their configurations of drivers.

\section{Deterministic scale-free networks}

\begin{figure}
\centering
\includegraphics[width=0.42\textwidth]{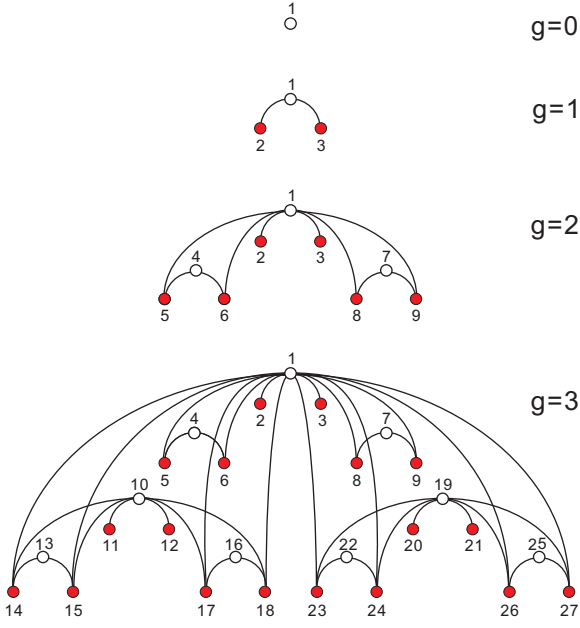}
\caption{The first four steps of construction of the DSFN model $D_{2,g}$. The model is a bipartite and sparse network when $g\rightarrow\infty$.}
\label{fig.1}
\end{figure}

The first DSFN was proposed by Barab\'{a}si et al.~\cite{BRV2001pa}, which is a self-similar hierarchical network model, built in an iterative way. It is illustrated in Figure~\ref{fig.1}, with empty circles and red filled circles respectively showing the hub and rim nodes, which has a bipartite structure. In this classic model, the total number of nodes and that of links (edges) in the $g$th step (generation) are counted as $N_{2,g}=3^{g}$ and $E_{2,g}=2(3^{g}-2^{g})$, respectively~\cite{IY2005pre}, implying that this network is sparse when $g$ is large enough. Thus, for the $g$th generation $D_{2,g}$ of the DSFN, the minimum number $N_{{D}}(D_{2,g})$ of drivers can be determined conveniently and analytically.

Let $A_{2,g}$ stand for the adjacency matrix of $D_{2,g}$. Using the numbering shown in Figure~\ref{fig.1}, the adjacency matrices are expressed by~\cite{IY2005pre} 
\begin{equation}
\label{A21.mat.}
A_{2,1}=
\left[\begin{array}{ccc}
0&1&1\\
1&0&0\\
1&0&0
\end{array}\right],
\end{equation}
\begin{equation}
\label{A22.mat.}
A_{2,2}=
\left[\begin{array}{ccccccccc}
0&1&1&0&1&1&0&1&1\\
1&0&0&&&&&&\\
1&0&0&&&&&&\\
0&&&0&1&1&&&\\
1&&&1&0&0&&&\\
1&&&1&0&0&&&\\
0&&&&&&0&1&1\\
1&&&&&&1&0&0\\
1&&&&&&1&0&0
\end{array}\right],
\end{equation}
and so forth. Here all blanks are zeros. Obviously, matrix $A_{2,1}$ can be transformed as: 
\begin{equation}
\label{A21.transform.}
A_{2,1}\xrightarrow[\mathscr{R}_{3}-\mathscr{R}_{2}]{\mathscr{C}_{3}-\mathscr{C}_{2}}
\left[\begin{array}{ccc}
0&1&0\\
1&0&0\\
0&0&0
\end{array}\right],
\end{equation}
and $A_{2,2}$ transformed as a block diagonal matrix: 
\begin{equation}
\label{A22.transform.}
A_{2,2}\rightarrow
\left[\begin{array}{ccc}
A_{2,1}&&\\
&A_{2,1}&\\
&&A_{2,1}
\end{array}\right],
\end{equation}
by operations of $\mathscr{C}_{9}-\mathscr{C}_{2}$, $\mathscr{C}_{8}-\mathscr{C}_{2}$, $\mathscr{C}_{6}-\mathscr{C}_{2}$, $\mathscr{C}_{5}-\mathscr{C}_{2}$, $\mathscr{R}_{9}-\mathscr{R}_{2}$, $\mathscr{R}_{8}-\mathscr{R}_{2}$, $\mathscr{R}_{6}-\mathscr{R}_{2}$, and $\mathscr{R}_{5}-\mathscr{R}_{2}$. In the same way, it can be proven that $A_{2,g}$ can also be transformed as a block diagonal matrix: 
\begin{equation}
\label{A2g.transform.}
A_{2,g}\rightarrow
\left[\begin{array}{ccc}
A_{2,g-1}&&\\
&A_{2,g-1}&\\
&&A_{2,g-1}
\end{array}\right].
\end{equation}
So, for any $g\geqslant 1$, we have $\text{rank}(A_{2,g})=3\text{rank}(A_{2,g-1})=\ldots=3^{g-1}\text{rank}(A_{2,1})=2\times 3^{g-1}$, which implies $\delta_{2,g}(0)=3^{g}-\text{rank}(A_{2,g})=3^{g-1}=N_{2,g}/3$. According to equations~(\ref{ND.delta}) and (\ref{ND.max.N.A}), the minimum number $N_{D}$ of drivers for $D_{2,g}$ is given by 
\begin{equation}
\label{ND.D2g}
N_{D}(D_{2,g})=\max\{1,3^{g}-\text{rank}(A_{2,g})\}=3^{g-1}.
\end{equation}
The controllability $n_{D}$ of the DSFN $D_{2,g}$ is then calculated as a constant: 
\begin{equation}
\label{nD.def.D2g}
n_{D}(D_{2,g})=\frac{N_{D}(D_{2,g})}{N_{2,g}}=\frac{3^{g-1}}{3^{g}}=\frac{1}{3}.
\end{equation}
From transformations (\ref{A21.transform.})--(\ref{A2g.transform.}), we can identify the redundant rows or drivers for the DSFN $D_{2,g}$ (see Tab.~\ref{tab.1}). For instance, from the transformed matrix (\ref{A21.transform.}) of $A_{2,1}$, we can determine the driver of $D_{2,1}$ to be node $3$ because the third row is redundant. Of course, we can also choose node $2$ as the driver of $D_{2,1}$ if we transform the second row of $A_{2,1}$ in (\ref{A21.mat.}) into a redundant row. Obviously, node $2$ is topologically equivalent to node $3$ because the topology of $D_{2,1}$ remains unchanged and $A_{2,1}$ is invariant under a $180^{\circ}$ rotation of $D_{2,1}$. Similarly in $D_{2,2}$, node $5$ is topologically equivalent to node $6$, and node $8$ to node $9$. Thus the minimum set of drivers for the DSFN is not unique. It is not difficult to obtain all possible minimum sets of drivers for the DSFN: sets $\{3\}$ and $\{2\}$ for $D_{2,1}$; sets $\{3,6,9\}$, $\{3,6,8\}$, $\{3,5,9\}$, $\{3,5,8\}$, $\{2,6,9\}$, $\{2,6,8\}$, $\{2,5,9\}$ and $\{2,5,8\}$ for $D_{2,2}$; and sets $\{1,2,\dots,3^{g}\}-\{3i+1,3i+k\,|\,i=0,1,\dots,3^{g-1}-1\}$ ($k\in\{2,3\}$) for $D_{2,g}$. For a specific generation $g$, these possible sets are topologically equivalent, so we only give one of them in Table~\ref{tab.1}.

\begin{table*}
\caption{The minimum number $N_{D}$ and configuration of drivers for DSFNs ($D_{2,g}$ and $D_{b,g}$) and Cayley trees ($C_{3,g}$ and $C_{b,g}$). The minimum set of drivers is not unique.}
\label{tab.1}
\begin{center}
\begin{tabular}{ccc}
\hline
Network & $N_{D}$ & Minimum set of drivers\\
\hline
$D_{2,1}$ & $1$ & $\{3\}$\\
$D_{2,g}$\ ($g\geqslant 1$) & $3^{g-1}$ & $\{3,6,\dots,3^{g}\}$\\[4pt]
$D_{b,1}$ & $b-1$ & $\{3,4,\dots,b+1\}$\\
$D_{b,g}$\ ($g\geqslant 1$) & $(b-1)(b+1)^{g-1}$ & $\{i(b+1)+3,i(b+1)+4,\dots,(i+1)(b+1)\,|\,i=0,1,\dots,(b+1)^{g-1}-1\}$\\[4pt]
$C_{3,1}$ & $2$ & $\{3,4\}$\\
$C_{3,g}$\ (odd $g\geqslant 1$) & $2^{g}$ 
& $\beta^{-}_{3,g}\cup\beta^{-}_{3,g-2}\cup\ldots\cup\beta^{-}_{3,1}$\\
$C_{3,g}$\ (even $g\geqslant 2$) & $2^{g}$ 
& $\beta^{-}_{3,g}\cup\beta^{-}_{3,g-2}\cup\ldots\cup\beta^{-}_{3,0}$\\[4pt]
$C_{b,1}$ & $b-1$ & $\{3,4,\dots,b+1\}$\\
$C_{b,g}$\ (odd $g\geqslant 1$) & $(b-1)^{g}$ 
& $\beta^{-}_{b,g}\cup\beta^{-}_{b,g-2}\cup\ldots\cup\beta^{-}_{b,1}$\\
$C_{b,g}$\ (even $g\geqslant 2$) & $(b-1)^{g}$ 
& $\beta^{-}_{b,g}\cup\beta^{-}_{b,g-2}\cup\ldots\cup\beta^{-}_{b,0}$\\
[2pt]\hline
\end{tabular}
\end{center}
\end{table*}

For the general case $D_{b,g}$ of the DSFN with the construction of one hub node generating $b$ ($b\geqslant 2$) rim nodes or subgraphs, the total number of nodes and that of edges in the $g$th generation are $N_{b,g}=(b+1)^{g}$ and $E_{b,g}=b((b+1)^{g}-b^{g})$, respectively. The above results for the classic DSFN $D_{2,g}$ can be easily generalized to $D_{b,g}$. We can obtain the minimum number of drivers for $D_{b,g}$ to be 
\begin{equation}
\label{ND.Dbg}
N_{D}(D_{b,g})=(b-1)(b+1)^{g-1},
\end{equation}
and the controllability of $D_{b,g}$ to be 
\begin{equation}
\label{nD.def.Dbg}
n_{D}(D_{b,g})=\frac{N_{D}(D_{b,g})}{N_{b,g}}=\frac{b-1}{b+1}.
\end{equation}
All possible minimum sets of drivers are: 
$\{1,2,\dots,b+1\}-\{1,k\}$ ($k\in\{2,3,\dots,b+1\}$) for $D_{b,1}$; 
$\{1,2,\dots,(b+1)^{2}\}-\{i(b+1)+1,i(b+1)+k\,|\,i=0,1,\dots,b\}$ 
($k\in\{2,3,\dots,b+1\}$) for $D_{b,2}$; and 
$\{1,2,\dots,(b+1)^{g}\}-\{i(b+1)+1,i(b+1)+k\,|\,i=0,1,\dots,(b+1)^{g-1}-1\}$ 
($k\in\{2,3,\dots,b+1\}$) for $D_{b,g}$. 
As these sets are topologically equivalent for a specific generation $g$, so we only list one of them in Table~\ref{tab.1} without loss of generality. For other self-similar DSFNs generated by iteration, their controllability and drivers can also be investigated similarly.

\section{Cayley trees}

\begin{figure}
\centering
\includegraphics[width=0.28\textwidth]{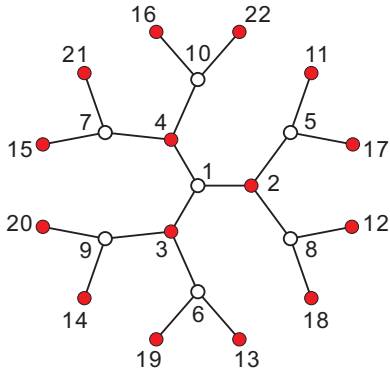}
\caption{The Cayley tree $C_{3,3}$. The network shows a bipartite structure.}
\label{fig.2}
\end{figure}

Cayley trees are a classic model for dendrimers~\cite{CC1997m,JWZ2013jcp}. Let $C_{b,g}$ ($b\geqslant 3$, $g\geqslant 0$) denote the Cayley trees after $g$ generations, which can be constructed in the following iterative fashion. Initially ($g=0$), $C_{b,0}$ is just a central node (the core). Next ($g=1$), $b$ nodes are generated connecting the central node to form $C_{b,1}$, with the $b$ single-degree nodes constituting the boundary nodes of $C_{b,1}$. For any $g\geqslant 2$, $C_{b,g}$ is built from $C_{b,g-1}$: for each boundary node of $C_{b,g-1}$, $b-1$ new nodes are created and linked to the boundary node. A specific Cayley tree $C_{3,3}$ is shown in Figure~\ref{fig.2}.

Let $N_{b}(i)$ denote the number of new nodes created at generation (iteration) $i$. It is easy to verify that 
\begin{equation}
\label{Nb.i.new}
N_{b}(i)=
\left\{\begin{array}{ll}
1,&i=0;\\
b(b-1)^{i-1},&i\geqslant 1.
\end{array}\right.
\end{equation}
Thus, the total number of nodes in $C_{b,g}$ (network size of $C_{b,g}$) is 
\begin{equation}
\label{Nb.g.total}
N_{b,g}=\sum\limits_{i=0}^{g} N_{b}(i)=\frac{b(b-1)^{g}-2}{b-2},
\end{equation}
and the total number of edges in $C_{b,g}$ is $E_{b,g}=N_{b,g}-1$. Both equations~(\ref{Nb.i.new}) and (\ref{Nb.g.total}) can be written recursively as ($g\geqslant 2$): $N_{b}(g)=(b-1)N_{b}(g-1)$, $N_{b,g}=N_{b,g-1}+(b-1)N_{b}(g-1)$. Obviously, Cayley trees are bipartite and sparse.

Here, we consider a specific Cayley tree with $b=3$, since for other values of $b$ the calculation and result are similar. Denote by $A_{3,g}$ the adjacency matrix of the $g$th generation Cayley tree $C_{3,g}$. Using the numbering given in Figure~\ref{fig.2}, we have 
\begin{equation}
\label{A31.transform.}
A_{3,1}=
\left[\begin{array}{cccc}
0&1&1&1\\
1&0&0&0\\
1&0&0&0\\
1&0&0&0
\end{array}\right]
\xrightarrow
[\begin{array}{c}
\mathscr{R}_{4}-\mathscr{R}_{2}\\
\mathscr{R}_{3}-\mathscr{R}_{2}
\end{array}]
{\begin{array}{c}
\mathscr{C}_{4}-\mathscr{C}_{2}\\
\mathscr{C}_{3}-\mathscr{C}_{2}
\end{array}}
\left[\begin{array}{cccc}
0&1&0&0\\
1&0&0&0\\
0&0&0&0\\
0&0&0&0
\end{array}\right].
\end{equation}
For $g\geqslant 2$, let $\alpha$ represent the set of nodes belonging to $C_{3,g-1}$, and $\beta$ the set of newly created nodes at the $g$th generation. Then, the $N_{3,g}\times N_{3,g}$ square matrix $A_{3,g}$ ($g\geqslant 2$) can be written in the following block form: 
\begin{equation}
\label{A3g.block form}
A_{3,g}=
\left[\begin{array}{cc}
A_{\alpha,\alpha}&A_{\alpha,\beta}\\
A_{\beta,\alpha}&A_{\beta,\beta}
\end{array}\right]=
\left[\begin{array}{cc}
A_{3,g-1}&A_{\alpha,\beta}\\
A_{\beta,\alpha}&O
\end{array}\right],
\end{equation}
where the matrix $A_{\alpha,\alpha}$ is just equal to $A_{3,g-1}$, $A_{\beta,\beta}$ is a zero matrix $O$ since there is no connection between any two nodes in $\beta$, and $A_{\alpha,\beta}$ ($=A_{\beta,\alpha}^\text{T}$) can be explained by zero and identity matrices as: 
\begin{equation}
\label{A.beta.alpha}
A_{\beta,\alpha}=
\left[\begin{array}{cc}
O_{N_{3}(g-1)\times N_{3,g-2}}&I_{N_{3}(g-1)}\\
O_{N_{3}(g-1)\times N_{3,g-2}}&I_{N_{3}(g-1)}
\end{array}\right].
\end{equation}
To find the drivers from $C_{3,g}$, reducing $A_{3,g}$ is needed. Let $\gamma$ represent the set $\{1,2,\dots,N_{3,g}\}$, and $a_{i,j}$ the element at row $i$ and column $j$ of matrix $A_{3,g}$. To begin with, performing on $A_{3,g}$ a series of elementary column transformations: 
$\mathscr{C}_{i}-a_{N_{3,g-2}+j,i}\mathscr{C}_{N_{3,g-1}+j}$ where $i\in\gamma-\{N_{3,g-1}+j\}$, $j\in\{1,2,\dots,N_{3}(g-1)\}$, we know that $A_{3,g}$ can be transformed as: 
\begin{equation}
\label{A3g.column transform.}
A_{3,g}\rightarrow
\left[\begin{array}{cccc}
*&*&O&O\\
O&O&I_{N_{3}(g-1)}&O\\
O&I_{N_{3}(g-1)}&O&O\\
O&I_{N_{3}(g-1)}&O&O
\end{array}\right],
\end{equation}
where $*$ denotes the entries that remain unchanged. Consequently, by a series of elementary row transformations similarly, $A_{3,g}$ can be further reduced as: 
\begin{equation}
\label{A3g.row transform.}
A_{3,g}\rightarrow
\left[\begin{array}{cccc}
*&O&O&O\\
O&O&I_{N_{3}(g-1)}&O\\
O&I_{N_{3}(g-1)}&O&O\\
O&O&O&O
\end{array}\right],
\end{equation}
where $*$, denoting the entries unchanged from $A_{3,g}$, is precisely the entries of $A_{3,g-2}$. So, for $g\geqslant 2$, $A_{3,g}$ can eventually be transformed as: 
\begin{equation}
\label{A3g.transform.}
A_{3,g}\rightarrow
\left[\begin{array}{cccc}
A_{3,g-2}&O&O&O\\
O&O&I_{N_{3}(g-1)}&O\\
O&I_{N_{3}(g-1)}&O&O\\
O&O&O&O
\end{array}\right].
\end{equation}
Thus, we have $\text{rank}(A_{3,g})=2N_{3}(g-1)+\text{rank}(A_{3,g-2})$.

Similarly, we can obtain 
\begin{equation}
\label{rank.Abg.g-2}
\text{rank}(A_{b,g})=2N_{b}(g-1)+\text{rank}(A_{b,g-2}).
\end{equation}
Note that $\text{rank}(A_{b,0})=0$ and $\text{rank}(A_{b,1})=2$. So, if $g$ ($\geqslant 2$) is even, then 
\begin{eqnarray}
\text{rank}(A_{b,g}) & = & 2N_{b}(g-1)+2N_{b}(g-3)+\dots   \nonumber \\
& & +2N_{b}(1)+\text{rank}(A_{b,0})  \nonumber \\
& = & \frac{2(b-1)^{g}-2}{b-2};  \label{rank Abg.even g}
\end{eqnarray}
if $g$ ($\geqslant 3$) is odd, then 
\begin{eqnarray}
\text{rank}(A_{b,g}) & = & 2N_{b}(g-1)+2N_{b}(g-3)+\dots   \nonumber \\
& & +2N_{b}(2)+\text{rank}(A_{b,1})  \nonumber \\
& = & \frac{2(b-1)^{g}-2}{b-2}.  \label{rank Abg.odd g}
\end{eqnarray}
Thus, for any $g\geqslant 0$, we always have 
\begin{equation}
\label{rank.Abg}
\text{rank}(A_{b,g})=\frac{2(b-1)^{g}-2}{b-2}.
\end{equation}
Therefore, the algebraic multiplicity of eigenvalue $0$ for $C_{b,g}$ can be deduced as 
\begin{equation}
\label{delta.g}
\delta_{b,g}(0)=N_{b,g}-\text{rank}(A_{b,g})=(b-1)^{g},
\end{equation}
which implies $\delta_{b,g}(0)>N_{b,g}/3$. According to equations~(\ref{ND.delta}) and (\ref{ND.max.N.A}), we obtain 
\begin{equation}
\label{ND.Cbg}
N_{D}(C_{b,g})=\delta_{b,g}(0)=(b-1)^{g}.
\end{equation}
The controllability $n_{D}$ of the Cayley tree $C_{b,g}$ can be given by 
\begin{equation}
\label{nD.def.Cbg}
n_{D}(C_{b,g})=\frac{N_{D}(C_{b,g})}{N_{b,g}}=\frac{(b-1)^{g}(b-2)}{b(b-1)^{g}-2}
\end{equation}
with its thermodynamic limit 
\begin{equation}
\label{lim.nD.def.Cbg}
\lim\limits_{g\rightarrow\infty}n_{D}(C_{b,g})=\lim\limits_{g\rightarrow\infty}\frac{(b-1)^{g}(b-2)}{b(b-1)^{g}-2}=\frac{b-2}{b}
\end{equation}
for any given $b$ ($\geqslant 3$).

\begin{figure}
\centering\
\includegraphics[width=0.44\textwidth]{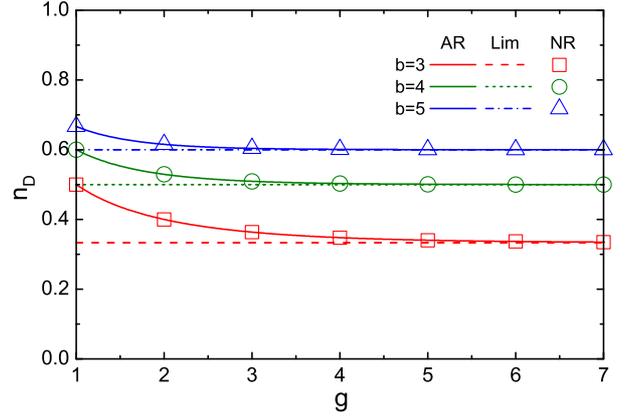}
\caption{Controllability measure $n_{D}$ of Cayley trees $C_{b,g}$ ($b=3,4,5$) as a function of generation $g$: AR denotes the analytical results predicted by equation~(\ref{nD.def.Cbg}), Lim the thermodynamic limit predicted by equation~(\ref{lim.nD.def.Cbg}) and NR the numerical results calculated by equation~(\ref{ND.max.delta}), respectively.}
\label{fig.3}
\end{figure}

In Figure~\ref{fig.3}, we show the analytical results of the controllability $n_{D}$ of Cayley trees $C_{b,g}$ with $b=3$, $4$ and $5$, respectively. We see that the controllability measure $n_{D}$ decreases monotonically and approaches rapidly its limit as $g$ increases, and the analytical results obtained from equation~(\ref{nD.def.Cbg}) are in exactly agreement with the numerical results based on equation~(\ref{ND.max.delta}). Here, when $g$ increases, although both the minimum number $N_{D}(C_{b,g})$ of drivers and the network size $N_{b,g}$ increase exponentially, the ratio of the increase of drivers to that of nodes is constant for two adjacent $g$ values: $\Delta N_{D}(C_{b,g})/\Delta N_{b,g}=(b-2)/b$, thus leads to the thermodynamic limit of $n_{D}$ as shown in equation~(\ref{lim.nD.def.Cbg}) and Figure~\ref{fig.3}.

For $C_{b,g}$ ($b\geqslant 3$, $g\geqslant 0$), let $\beta_{b,j}=\{N_{b,j-1}+1,N_{b,j-1}+2,\dots,N_{b,j}\}$ ($j\leqslant g$) denote an ordered set of newly created nodes at the $j$th generation. Obviously, the number of elements of $\beta_{b,j}$ is $|\beta_{b,j}|=N_{b}(j)$. By removing the first $N_{b}(j-1)$ elements from $\beta_{b,j}$, we can get another set denoted by $\beta^{-}_{b,j}=\{N_{b,j-1}+N_{b}(j-1)+1,N_{b,j-1}+N_{b}(j-1)+2,\dots,N_{b,j}\}$ with $(b-2)N_{b}(j-1)$ elements. For example, $\beta^{-}_{3,0}=\beta^{-}_{b,0}=\{1\}$, $\beta^{-}_{3,1}=\{2,3,4\}-\{2\}=\{3,4\}$, and $\beta^{-}_{b,1}=\{3,4,\dots,b+1\}$. From transformations (\ref{A31.transform.}) and (\ref{A3g.transform.}), we can thus identify the redundant rows or drivers for (dendrimers modeled by) Cayley trees (see Tab.~\ref{tab.1}). Here, the choice of $\beta^{-}_{b,j}$ is not unique: one can also, more generally, obtain $\beta^{-}_{b,j}$ by removing other elements from $\beta_{b,j}$, such as 
$\beta^{-}_{b,0}=\{1\}$, 
$\beta^{-}_{b,1}=\beta_{b,1}-\{s_{1}\}$ ($s_{1}\in\beta_{b,1}$), 
$\beta^{-}_{b,j}=\beta_{b,j}-\{s_{1},s_{2},\ldots,s_{N_{b}(j-1)}\}$ ($j\geqslant 2$) 
where $s_{k}\in\{N_{b,j-1}+iN_{b}(j-1)+k\,|\,i=0,1,\ldots,b-2\}$ ($k=1,2,\ldots,N_{b}(j-1)$); 
from this choice, we have the expressions 
$\beta^{-}_{b,g}\cup\beta^{-}_{b,g-2}\cup\ldots\cup\beta^{-}_{b,1}$ (for odd $g\geqslant 1$) 
and 
$\beta^{-}_{b,g}\cup\beta^{-}_{b,g-2}\cup\ldots\cup\beta^{-}_{b,0}$ (for even $g\geqslant 2$), 
giving all possible minimum sets of drivers for Cayley trees $C_{b,g}$.

\section{Discussion and conclusions}

From Table~\ref{tab.1}, we can see that, although these self-similar networks analyzed are all constructed by iteration, their exact controllability $n_{D}$ can be quite different. The controllability may vary with parameters in some situations such as in Cayley trees (see Eq.~(\ref{nD.def.Cbg}) and Fig.~\ref{fig.3}) or be just constants such as in DSFNs (see Eqs.~(\ref{nD.def.D2g}) and (\ref{nD.def.Dbg})).

\begin{figure}
\centering
\includegraphics[width=0.44\textwidth]{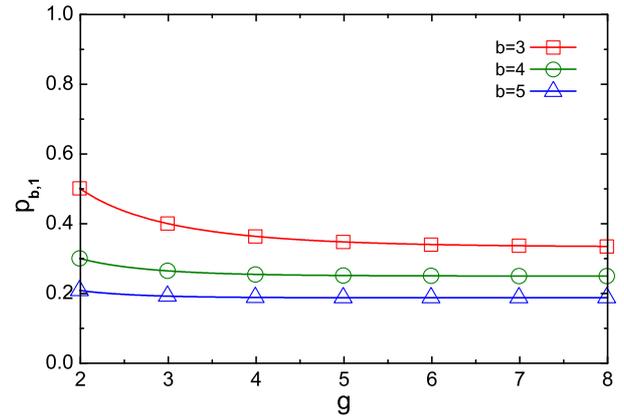}
\caption{Degree preference index $p_{b,1}(g)$ of Cayley trees $C_{b,g}$ ($b=3,4,5$) measured by equation~(\ref{deg.prefer.}).}
\label{fig.4}
\end{figure}

The configuration of drivers is related to the degree distribution. For DSFNs $D_{b,g}$ ($b\geqslant 2$), the drivers all come from the rim nodes, which implies that the drivers are likely located on the low-degree nodes. For Cayley trees $C_{b,g}$ ($b\geqslant 3$, $g\geqslant 2$), the nodes can be divided into two parts: $\alpha_{b,g-1}=\{1,2,\dots,N_{b,g-1}\}$ (the set of nodes of $C_{b,g-1}$) and $\beta_{b,g}=\{N_{b,g-1}+1,N_{b,g-1}+2,\dots,N_{b,g}\}$; obviously, $\alpha_{b,g}=\alpha_{b,g-1}\cup\beta_{b,g}$. We know that the degree of any node in $\beta_{b,g}$ is $1$, and the proportion of drivers in $\beta_{b,g}$ is $n_{D,1}(\beta_{b,g})=|\beta^{-}_{b,g}|/|\beta_{b,g}|=(b-2)/(b-1)$. While the degree of every node in $\alpha_{b,g-1}$ is $b$ ($\geqslant 3$), and correspondingly, the proportion of drivers in $\alpha_{b,g-1}$ is $n_{D,b}(\alpha_{b,g-1})=((b-1)^{g}-|\beta^{-}_{b,g}|)/|\alpha_{b,g-1}|=(b-1)^{g-2}(b-2)/(b(b-1)^{g-1}-2)$. To compare them, we introduce a degree preference index defined as their ratio: 
\begin{equation}
\label{deg.prefer.}
p_{b,1}(g)=\frac{n_{D,b}(\alpha_{b,g-1})}{n_{D,1}(\beta_{b,g})}=\frac{(b-1)^{g-1}}{b(b-1)^{g-1}-2}\leqslant\frac{1}{b-1},
\end{equation}
with the limit $\lim_{g\rightarrow\infty}p_{b,1}(g)=1/b$. Obviously, $p_{b,1}(g)$ is always smaller than $1$, which means that the proportion of drivers among high-degree nodes, $n_{D,b}$, is lower (see Fig.~\ref{fig.4}). Moreover, from the perspective of the transformations of the coupling matrix, the rows and columns of low-degree nodes are more easily eliminated. Therefore, the drivers in both models tend to avoid the high-degree nodes (or hubs), which is in agreement with reference~\cite{LSB2011n}.

In fact, our study can be extended to DSFNs and Cayley trees with arbitrary nonzero link weights. It is worth noting that the results remain exactly the same because the weight values do not affect the structure of adjacency matrices and elementary transformations are also similar. That is, the exact controllability and the distribution of drivers are independent of link weights in these two types of networks. In other words, they are robust to the variation of link weights, showing a characteristic of strong structural controllability~\cite{L1974tac,CM2013acc}. A possible reason for this robustness lies in that the self-similar structures of adjacency matrices are always preserved under elementary transformations. Such a robustness would be of significance, as the link weights for most real networks are either unknown or known only approximately.

All in all, the controllability of self-similar networks can be predicted by applying the exact controllability theory. In particular, for undirected bipartite networks, we provide a sufficient condition~(\ref{suff.cond.}) for the expression (Eq.~(\ref{ND.max.N.A})) of the minimum number $N_{D}$ of drivers. So, the controllability $n_{D}$ can be completely determined by the rank of the coupling matrix, which can be analytically derived by elementary matrix transformations because of the self-similarity of the network. Furthermore, using these transformations properly, we can identify the drivers (driver nodes) and reveal their distribution characteristics. In this paper, two prototypical self-similar bipartite networks, including DSFNs and Cayley trees, have been explored to validate our analytical results. From our research experience, exploring structural features of coupling matrices by elementary transformations, which can determine not only the controllability but also the drivers' distributions, will contribute to a deeper understanding for the control of real systems. Even for large systems, the self-similarity of networks also enables us to obtain analytical expressions of their controllability and configurations of their drivers. However, there are still some unclear issues surrounding the control in these networks with self-similarity. It would be important and interesting to further investigate the control energy~\cite{YRL2012prl} and target control~\cite{GLDS2014nc} of such networks.

\bigskip

\noindent\small{This work was supported by the National Natural Science Foundation of China (NSFC) (Grant No. 11365023). M. Xu was supported by the Scientific and Technological Fund of Guizhou Province (Grant No. J-2013-2260), the Joint Fund of Guizhou Province (Grant Nos. LH-2014-7231 and J-LKK-2013-31), and the Natural Science Research Project of the Department of Education of Guizhou Province (Grant No. KY-2013-185). The authors would like to thank Prof. S.-L. Peng for his helpful discussions, and the two anonymous referees for their constructive comments and suggestions. M. Xu and K.-F. Cao conceived and designed the study. M. Xu, C.-Y. Xu, H. Wang and C.-Z. Deng performed the analytical and numerical calculations. M. Xu drafted the manuscript. K.-F. Cao supervised the research, and revised and finalized the manuscript with M. Xu and C.-Y. Xu.}

%
%

\end{document}